\documentclass[12pt]{article}
\usepackage{amssymb,amsmath}
\usepackage{blkarray}
\title{The algorithmic randomness of quantum measurements}
\author{Mohammad Shahbazi\footnote{Department of Mathematics, School of Mathematics, Statistics and Computer Science, College of Science, University of Tehran, Tehran, Iran} \\ mshahbazi@ut.ac.ir }
\date{17 Jan 2017}

\begin{document}
\maketitle

\section*{abstract}

This paper is a comment on the paper ``Quantum Mechanics and Algorithmic Randomness'' was written by Ulvi Yurtsever \cite{Yurtsever} and the briefly explanation of the algorithmic randomness of quantum measurements results. 

There are differences between the computability of probability sources, ( which means there is an algorithm that can define the way that random process or probability source generates the numbers ) and the algorithmic randomness of the sequences or strings which are produced by a source.
We may have the source without a computable algorithm for that but it can produce compressible or incompressible strings.
For example, so far there is no computable algorithm that can define the abstract meaning of randomness even the easiest one, Bernoulli probability distribution. Historically and philosophically there many scientist believe the existence of the algorithm for a random process is a contradiction because in their opinion, in the definition of a random variable, implicitly assumed that there is no reason for the happening of an event and we just know the probabilities. There is however no need to enter into this matter here. As in the paper mentioned, all the algorithms for simulating a random process try to pass the statistical tests and be close to the abstract meaning of it.

\section{Introduction}

Before quantum mechanics be introduced, there was this approach that there is no real randomness in the real world and the probability models are just theoretical concepts and the randomness that will defined in the real world, as it was in the mentioned paper too, it is just pseudo-randomness which means they are simulation of the abstract meaning of randomness and they have been introduced to have similar property as the abstract definition\cite{Gillies}. Because in the definition of a random probability we just know with some probabilities some events happen and there is no reason for them. Defining every reason for that change the definition of randomness which this means there is no algorithm that can predict and simulate the outcomes of a probability source. By this abstract definition of each random variable or probability source is not computable, this means there is no algorithm for defining them.

After quantum mechanics as you can see \cite{Heisenberg}, the results of quantum measurement define exactly similar to the abstract meaning of probability theory, and they suppose there is no reason that why this result appear, it is important that they emphasize, there is no reason and it is different from, there can be a reason and we have lack of knowledge.

Due to this definition the quantum randomness is completely correspond to the abstract mathematical definition of probability randomness and it not computable too.

As you can see in the definition of the Bernoulli probability source there is no reason or algorithm that why the probability source produce zero or one and it is just happens with probability p.
But the strings and the sequences are produced by these sources can be algorithmically random or can be compressible.  

With above definitions, theoretically there is no differences between Bernoulli sources and  measurements on qubits or $2\times 2$ density metrices, but practically similar to \cite{Calude}, it is possible to find difference between classical sequences and quantum measurements results. 

\section{Preliminaries}
In this section we will discuss some basic concepts about quantum models and quantum measurements that we need here, then if you are familiar with these topics, you can jump to the conclusions, at the section ``About this article''. 
\subsection{Quantum sequences}

For each density matrix $\rho\in {M_2}(\mathbb{C})$, we can perform various measurements  corresponds to a family of operators $ \Pi = \lbrace\pi_{0},\pi_{1}\rbrace \in {M_2}(\mathbb{C})$ satisfying the following:
\begin{equation} \label{eq:s1}
\pi_{k} = \pi^{*}_{k}, \quad \pi_{k}^{2}=\pi_{k} ,\quad k\neq l \Rightarrow \pi_{k} \pi_{l} =0,\quad \pi_{0}+\pi_{1}=I
\end{equation}

density operators play the role of probability distributions in the ordinary probability theory.\cite{Hayashi}

\textbf{Property 1.} When we perform the measurement $\Pi$ under the state $\rho$ , the probability of obtaining the outcome $\pi_k$, say $\mathbb{P}(k\vert\Pi)$, is given by $\mathbb{P}(k\vert\Pi)=Tr(\rho\pi_k)$.

\textbf{Property 2.} The quantum model of performing measurement $ \Pi = \lbrace\pi_{0}=\vert 0\rangle\langle 0 \vert,\pi_{1}=\vert 1 \rangle\langle 1 \vert\rbrace \in {M_2}(\mathbb{C})$ on the quantum state
$\rho = 
\frac{1}{2}\begin{bmatrix}
1 & 0 \\
0 & 1 \\
\end{bmatrix},
$
are exactly the same as the classical model of Bernoulli distribution with parameter 1/2, which means in both model just two events can happen and each events happen with probability 1/2 and there is no different from doing one measurement on the state $\rho$ and one realization of a Bernoulli distribution.

\textbf{Property 3.} If $x_1 x_2 \cdots x_n$ be a string with length n and $x_{i}\in \lbrace 0 , 1\rbrace $. If this string generate from a Bernoulli source ( which means $x_1 x_2 \cdots x_n$ are i.i.d with Bernoulli distribution) there can be $2^n$ different strings with length n and all with probability $\frac{1}{2^n}$ and similarly this string with the same probability distribution for each case can be the results of the quantum measurements.
Generally, Let n be the sample size or the length of the string for the string $x_1 x_2 \cdots x_n$ we define $D=\vert x_{1}x_{2}\cdots x_{n}\rangle$, and $x_{i}\in \lbrace 0 , 1\rbrace $. The element of D is a result of following quantum measurements.

Let $ \Pi = \lbrace\pi_{0},\pi_{1}\rbrace \in {M_2}(\mathbb{C})$ satisfying the following:
\begin{equation} \label{eq:s1}
\pi_{k} = \pi^{*}_{k}, \quad \pi_{k}^{2}=\pi_{k} ,\quad k\neq l \Rightarrow \pi_{k} \pi_{l} =0,\quad \pi_{0}+\pi_{1}=I
\end{equation}
Let us denote $\rho^{\otimes(n)}:=\underbrace{\rho\otimes\rho\otimes\cdots\otimes\rho}_n$,
$ \Pi^{\otimes(n)} = \lbrace \pi_{i_1}\otimes\pi_{i_2}\otimes\cdots\pi_{i_n}\vert \pi_{i_j}\in \lbrace\pi_{1},\pi_{2}\rbrace\rbrace$

$$D=\vert x_{1}x_{2}\cdots x_{n}\rangle=  \frac{\pi_{x_1}\otimes\pi_{x_2}\otimes\cdots\pi_{x_n}( \rho^{\otimes(n)}) \pi_{x_1}\otimes\pi_{x_2}\otimes\cdots\pi_{x_n}}{Tr(\rho\pi_{x_0})Tr(\rho\pi_{x_1})\cdots Tr(\rho\pi_{x_n})}$$
$$ \frac{(\pi_{x_1} \rho\pi_{x_1}) \otimes(\pi_{x_2} \rho\pi_{x_2})\otimes\cdots \otimes(\pi_{x_n} \rho\pi_{x_n})}{Tr(\rho\pi_{x_0})Tr(\rho\pi_{x_1})\cdots Tr(\rho\pi_{x_n})}$$
with probability $\mathbb{P}(D)=Tr(\rho\pi_{x_0})Tr(\rho\pi_{x_1})\cdots Tr(\rho\pi_{x_n}).$

\textbf{Property 4.} The quantum strings that are generated by doing a measurement on density metrics $\rho\in{M_2}(\mathbb{C})$ and classical strings are generated by Bernoulli source, are theoretically the same.
Although practically the strings are produced by computers or flipping a real coin can be different from the quantum strings, because they are some simulations of Bernoulli distribution, and they just trying to be close to the abstract definition of Bernoulli distribution.

\textbf{Property 5.} As we defined above, The infinite sequence is generated by both of the models (quantum measurements and classical Bernoulli distribution) according to the brudno's theorem are algorithmically random.

As we defined, the sequence has been produced in quantum source or measurement of quantum system, can be similar Bernoulli distribution or Bernoulli source. According to the Heisnberg definition there is no reason for the results of quantum measurement. Then it is obvious that the quantum measurement is Kolmogorov randomness. But the sequence that will be produced by it can be compressible or incompressible.

\subsection{Quantum measurements on the entangled states }

\textbf{Remark 1.} Let $\vert \psi \rangle=\frac{\vert 00 \rangle+\vert 11 \rangle}{\sqrt{2}}\in \mathbb{H}_{A}\otimes\mathbb{H}_{B}$, be a shared state between Alice and Bob.
When Bob wants to do  an arbitrary measurement with the operators $ \lbrace M_{\alpha}, M_{\beta}\rbrace$, then the measurement operators on the whole system will be $ \lbrace I\otimes M_{\alpha},I\otimes M_{\beta}\rbrace$, then
\begin{itemize}
\item We observe $\alpha$ with probability $\mathbb{P}(\alpha)=\langle\psi\vert I\otimes M^{\dagger}_{\alpha}M_{\alpha}\vert\psi\rangle$, the original quibt $\psi$ will collapse to $\psi_{\alpha}:=\frac{I\otimes M_{\alpha}\vert\psi\rangle}{\Vert I\otimes M_{\alpha}\vert\psi\rangle\Vert}$, where $\Vert I\otimes M_{\alpha}\vert\psi\rangle\Vert^2=\mathbb{P}(\alpha).$
\item We observe $\beta$ with probability $\mathbb{P}(\beta)=\langle\psi\vert I\otimes M^{\dagger}_{\beta}M_{\beta}\vert\psi\rangle$ and the original quibt $\psi$ will collapse to $\psi_{\beta}:=\frac{M_{\beta}\psi}{\Vert I\otimes M_{\beta}\vert\psi\rangle\Vert}$, where $\Vert I\otimes M_{\beta}\vert\psi\rangle\Vert^2=\mathbb{P}(\beta).$
\end{itemize}

$\hfill\blacksquare$

\textbf{Remark 2.} Similarly, here we can just do a measurement on the first system, we should choose the measurement as  $\lbrace \vert N\rangle\langle N\vert \otimes I,\vert N^{\prime}\rangle\langle N^{\prime}\vert\otimes I\rbrace$. Because in the mentioned paper, Alice do the measurement in the computational basis, so we do our measurement on the first system, before Bob's measurement, with these operators $ \lbrace \vert 0\rangle\langle 0\vert \otimes I,\vert 1\rangle\langle 1\vert\otimes I\rbrace$, on the qubit $\psi$, 

\begin{itemize}
\item We observe $0$ with probability $\mathbb{P}(0)=\langle\psi\vert \vert 0\rangle\langle 0\vert \otimes I\vert\psi\rangle=1/2$ and the original quibt $\psi$ will collapse to $\psi_{0}:=\frac{\vert 0\rangle\langle 0\vert \otimes I\vert\psi\rangle}{\Vert \vert 0\rangle\langle 0\vert \otimes I\vert\psi\rangle\Vert}=\vert 00 \rangle$.
\item We observe $1$ with probability $\mathbb{P}(1)=\langle\psi\vert \vert 1\rangle\langle 1\vert \otimes I\vert\psi\rangle=1/2$ and the original quibt $\psi$ will collapse to $\psi_{1}:=\frac{\vert 1\rangle\langle 1\vert \otimes I\vert\psi\rangle}{\Vert \vert 1\rangle\langle 1\vert \otimes I\vert\psi\rangle\Vert}=\vert 11 \rangle$.
\end{itemize}

$\hfill\blacksquare$

\textbf{Remark 3.} Now, what happen if Alice does the measurement $ \lbrace \vert 0\rangle\langle 0\vert \otimes I,\vert 1\rangle\langle 1\vert\otimes I\rbrace$, after Bob's arbitrary measurement $ \lbrace I\otimes M_{\alpha},I\otimes M_{\beta}\rbrace$,

Let $\mathbb{P}(0\vert \alpha)=$ The probability of Alice observes 0 if Bob observed $\alpha$ in the second measurement. Then
\begin{equation*}
\begin{split}
\mathbb{P}(0\vert \alpha)&=\langle\psi_{\alpha}\vert \Big(\vert 0\rangle\langle 0\vert \otimes I\Big)\vert\psi_{\alpha}\rangle \\
&=\frac{\langle\psi\vert I\otimes M^{\dagger}_{\alpha}}{\Vert I\otimes M_{\alpha}\vert\psi\rangle\Vert}\Big(\vert 0\rangle\langle 0\vert \otimes I\Big) \frac{I\otimes M_{\alpha}\vert\psi\rangle}{\Vert I\otimes M_{\alpha}\vert\psi\rangle\Vert} \\
 &=\frac{\langle\psi\vert \vert 0\rangle\langle 0\vert\otimes M^{\dagger}_{\alpha}M_{\alpha}\vert\psi\rangle}{\Vert I\otimes M_{\alpha}\vert\psi\rangle\Vert^2}\\
 &=\frac{ \langle 0\vert 0\rangle \langle 0\vert M^{\dagger}_{\alpha}M_{\alpha}\vert 0\rangle}{2\Vert I\otimes M_{\alpha}\vert\psi\rangle\Vert^2}\\
 &=\frac{  \langle 0\vert M^{\dagger}_{\alpha}M_{\alpha}\vert 0\rangle}{2\Vert I\otimes M_{\alpha}\vert\psi\rangle\Vert^2}\\
\end{split}
\end{equation*}

Also $\mathbb{P}(1\vert \beta)=$ The probability of Alice observes 1 if Bob observed $\beta$ in the second measurement. Then,
\begin{equation*}
\begin{split}
\mathbb{P}(1\vert \beta)&=\langle\psi_{\beta}\vert\Big( \vert 1\rangle\langle 1\vert \otimes I\Big)\vert\psi_{\beta}\rangle \\
&=\frac{\langle\psi\vert I\otimes M^{\dagger}_{\beta}}{\Vert M_{\beta}\psi\Vert}\Big(\vert 1\rangle\langle 1\vert \otimes I\Big) \frac{I\otimes M_{\beta}\vert\psi\rangle}{\Vert I\otimes M_{\beta}\vert\psi\rangle\Vert} \\
 &=\frac{\langle\psi\vert \vert 1\rangle\langle 1\vert\otimes M^{\dagger}_{\beta}M_{\beta}\vert\psi\rangle}{\Vert I\otimes M_{\beta}\vert\psi\rangle\Vert^2}\\
 &=\frac{ \langle 1\vert 1\rangle \langle 1\vert M^{\dagger}_{\beta}M_{\beta}\vert 1\rangle}{2\Vert I\otimes M_{\beta}\vert\psi\rangle\Vert^2}\\
 &=\frac{  \langle 1\vert M^{\dagger}_{\beta}M_{\beta}\vert 1\rangle}{2\Vert I\otimes M_{\beta}\vert\psi\rangle\Vert^2} \\
\end{split}
\end{equation*}

with the \textit{law of total probability} if ${B_n, n=1,2,3,...}$ is a set of pairwise disjoint events whose union is the entire sample space, then for any event A of the same probability space:
$\mathbb{P}(A)=\sum_n \mathbb{P}(A\cap B_n)$ or equivalently
$$\mathbb{P}(A)=\sum_n \mathbb{P}(A\vert B_n)\mathbb{P}(B_n)$$

then in this case:

if $\mathbb{P}(0)$ = the probability that Alice observes 0 if Bob did an arbitrary measurement on the second system
then

\begin{equation*}
\begin{split}
\mathbb{P}(0)&=\mathbb{P}(0\vert \alpha)\mathbb{P}(\alpha)+\mathbb{P}(0\vert \beta)\mathbb{P}(\beta) \\
&= \frac{  \langle 0\vert M^{\dagger}_{\alpha}M_{\alpha}\vert 0\rangle}{2\Vert I\otimes M_{\alpha}\vert\psi\rangle\Vert^2}\times \Vert I\otimes M_{\alpha}\vert\psi\rangle\Vert^2 + \frac{  \langle 0\vert M^{\dagger}_{\beta}M_{\beta}\vert 0\rangle}{2\Vert I\otimes M_{\beta}\vert\psi\rangle\Vert^2}\times \Vert I\otimes M_{\beta}\vert\psi\rangle\Vert^2\\
 &=\frac{  \langle 0\vert M^{\dagger}_{\alpha}M_{\alpha}\vert 0\rangle}{2} + \frac{  \langle 0\vert M^{\dagger}_{\beta}M_{\beta}\vert 0\rangle}{2}\\
 &=\frac{  \langle 0\vert M^{\dagger}_{\alpha}M_{\alpha}+M^{\dagger}_{\beta}M_{\beta}\vert 0\rangle}{2}\\
 &=\frac{  \langle 0\vert I\vert 0\rangle}{2}\\
 &=\frac{ 1}{2}\\
\end{split}
\end{equation*}

and similarly $$\mathbb{P}(1)=\frac{ 1}{2}
$$

$\hfill\blacksquare$

\textbf{Remark 4.} Suppose $\lbrace L_l\rbrace$ and $\lbrace M_m\rbrace$ are two sets of measurement operators. Then a measurement defined by the measurement operators $\lbrace L_l\rbrace$ followed by a measurement defined by the measurement operators $\lbrace M_m\rbrace$ is physically equivalent to a single measurement defined by measurement operators $\lbrace N_{lm}\rbrace$ with the representation $N_{ml} \equiv M_m L_l$.\cite{Nielsen}

\textbf{Remark 5.} Let $${\vert\psi\rangle}^{\otimes(N)}:=\underbrace{{\vert\psi\rangle}\otimes{\vert\psi\rangle}\otimes\cdots\otimes{\vert\psi\rangle}}_N$$ be the quantum system of \textit{N} shared state between Alice and Bob where $\vert \psi \rangle=\dfrac{\vert 00 \rangle+\vert 11 \rangle}{\sqrt{2}}\in \mathbb{H}_{A}\otimes\mathbb{H}_{B}$ be a shared state between Alice and Bob.
 Bob does an arbitrary measurement on the \textit{j-th} state with the operators $ \lbrace \pi_{\alpha_j}, \pi_{\beta_j}\rbrace$, or similar Remark 2 does nothing and Alice performing measurement by operators $ \lbrace\vert 0\rangle\langle 0 \vert,\vert 1 \rangle\langle 1 \vert\rbrace$
then the measurement operators on the whole system will be

\begin{equation*}
\begin{split}
&\Pi^{\otimes(N)} =\bigg\lbrace \big(\pi_{i_1}\otimes\vert K_1\rangle\langle K_1\vert\big)\otimes
\big(\pi_{i_2}\otimes\vert K_2\rangle\langle K_2\vert\big)\otimes\cdots\big(\pi_{i_N}\otimes\vert K_N\rangle\langle K_N\vert\big)\bigg\vert\\
 &\pi_{i_j}\in \lbrace \pi_{\alpha_j}, \pi_{\beta_j}\rbrace, K_{i}\in \lbrace\vert 0\rangle\langle 0\vert,\vert 1\rangle\langle 1\vert\rbrace\bigg\rbrace 
\end{split}
\end{equation*} 
(Note: if Bob does more than one measurement as you can see in Remark 4, we can assume all of them as one measurement.)

Then according to the Property 3 and Remark 3, Alice observe each sequence $k_1 k_2 \cdots k_N$ where $k_j \in \lbrace 0,1 \rbrace$ with probability $\dfrac{1}{2^N}$, in other words 
 $$\mathbb{P}(\vert k_1 k_2 \cdots k_N\rangle)=\dfrac{1}{2^N}\qquad \forall k_j \in \lbrace 0,1 \rbrace.$$

\textbf{Remark 6.} According to the above discussions, the conclusion of these three following events are the same and \textit{in each case the probability of results, zero or one, comes from the Bernoulli distribution, which means Alice observes 0 with probability 1/2 and observes 1 with probability 1/2 and it will not change anything.}

\begin{enumerate}
\item If Bob does a measurement on the second system and then Alice does a measurement in the first system.
\item If Bob does many measurements (more than one) on the second system and then Alice does a measurement in the first system.
\item If Alice just does a measurement in the first system and Bob does nothing.
\end{enumerate}

$\hfill\blacksquare$

In every time, when we perform a measurement on one part of the entangled system, the probability of getting a specific result is completely independent from the measurement or unitary operators acting on the second part. As we showed in our case when we are doing the measurement in the computational basis on the state $\psi$, we always deal with a Bernoulli probability distribution with parameter p=1/2, then the result always comes from this Bernoulli distribution.

\section{About the article}
First, in the mentioned paper, we have many references to a manuscripts that doesn't have published yet.

Second, in the mentioned paper the writer defined a channel  for proving his theory and he said, \textit{``Here is how the construction of this communications channel might proceed: First, Alice and Bob agree at the outset that Alice
should interpret any compressible N-bit block in her string as a 0-bit, and any incompressible block as a 1-bit,.... \textbf{Now, to send a 0-bit to Alice, Bob does nothing} (i.e., keeps his spin-measurement axis unchanged, pointing along the original predetermined direction). \textbf{To send a 1-bit},..., Bob prepares a sequence of N random measurement directions using T as a random-number generator, and performs his next N measurements along the successive orthonormal bases associated with the successive directions from this random sequence. This procedure of scrambling with the random template T guarantees that Bob’s modified N-bit long string of quantum measurements is almost surely p-incompressible.'' }

and he proceeded to say,
\textit{``The crucial requirement for Bob’s choice of N measurements is that it should be guaranteed to be free of any regularities that might be present in the N-bit block (representing the 1-bit) he wishes to send Alice. His scrambling procedure is designed to destroy any nascent correlations that might exist in the original bit stream of measurements.''}

he did not prove that with this channel has a capacity more than zero and the results of using this channel is algorithmic randomness and by mistake, with the above descriptions he just explained why he thought  that it should have this properties. 

Then he concluded because of the channel he defined, we can send a message  with entangled quibts then it contradict with relativistic causality so the quantum randomness should be algorithmically random.

\begin{enumerate}
\item If Bob do every measurement in one entangled state $\psi$ similar to Remark 3, then changing Bob's measurements has no effect on the correlation or on the results that Alice will observe and the results come i.i.d., from a Bernoulli distribution (in contrast to the writer's speculation). Then as we discussed in Remark 5 and 6, and Property 4, it will not change the results that Alice will observe.
\item I am sure that the writer did not intend to do all different random measurements in one entangled measurement, but just to be clear, if he intended, it isn't helpful because with the first measurement that Alice does, the entangled state will collapse and the state of the system will be separable and the other Bob's measurements will not change the Alice state at all. 
\end{enumerate}

As you can see in the definition of channel if Bob wants to send zero he does nothing, he does not change the measurement basis, (he can do the measurement in this basis or he can do nothing) and if he wants to send 1 with the way was explained, he changes his measurement basis and does the measurement, and the writer think this randomly basis choosing can make an incompressible string for Alice, and he concluded ``\textit{This procedure of scrambling with the random template T guarantees that Bob's modified N-bit long string of quantum measurements is almost surely p-incompressible.}'' but as we mentioned above in Remark 5 and Remark 6, if Bob does nothing or chooses a basis randomly and does a measurement, Alice always deals with a Bernoulli probability distribution with parameter p=1/2, then the results always come from this Bernoulli distribution.
All the string with the length N will appear with probability $\frac{1}{2^N}$.
In both case (Bob does nothing, or act according his procedure) the string with N zeros comes exactly with probability $\frac{1}{2^N}$ similar to an incompressible string with length N can appear.
In the other words, Bob can not affect the Alice's strings, either he can not made the Alice's string become incompressible or he can not make it compressible, So his channel is useless.


\begin{thebibliography}{20}	

\bibitem{Yurtsever} U.Yurtsever. Quantum Mechanics and Algorithmic Randomness, arXiv:quant-ph/9806059. or Complexity journal, September 2000, DOI: 10.1002/1099-0526(200009/10)6:1<27::AID-CPLX1004>3.0.CO;2-R 

\bibitem{Nielsen} M. A. Nielsen and I. L. Chuang. Quantum Computation and Quantum Information, Cambridge University Press, 2010.

\bibitem{Hayashi} M. Hayashi. Asymptotic theory of quantum statistical inference, World Scientific Publishing Company, 2005. 

\bibitem{Gillies} Donald A. Gillies, Philosophical Theories of Probability, Philosophical Issues in Science, Routledge, 2000.

\bibitem{Heisenberg} W. Heisenberg, Der Teil Und Das Ganze: Gesprache Im Umbkreis Der Atomphysik, Piper Verlag Gmbh, 2002.

\bibitem{Calude} C. S. Calude, M. J. Dinneen, Monica Dumitrescu, K. Svozil. Experimental evidence of quantum randomness incomputability, Physical Review A, 82, 022102 (2010), 1-8.



\end{thebibliography}
\end{document}